\documentstyle[12pt]{article}
\def\pc{\,{\rm pc}}
\def\kpc{\,{\rm kpc}}
\def\Mpc{\,{\rm Mpc}}

\def\eV{\,{\rm eV}}
\def\MeV{\,{\rm MeV}}

\def\TeV{\,{\rm TeV}}
\def\G{\,{\rm G}}
\def\muG{\,\mu{\rm G}}
\def\gcm2{\,{\rm g\,cm}^{-2}}
\def\km{\,{\rm km}}
\def\sec{{\rm s}}
\def\la{\mathrel{\mathpalette\fun <}}
\def\ga{\mathrel{\mathpalette\fun >}}
\def\fun#1#2{\lower3.6pt\vbox{\baselineskip0pt\lineskip.9pt
  \ialign{$\mathsurround=0pt#1\hfil##\hfil$\crcr#2\crcr\sim\crcr}}}
\begin{document}
\pagestyle{empty}
\begin{center}
\vspace*{3cm}
{\Large \bf ON THE ORIGIN OF}\\
\bigskip
{\Large \bf HIGHEST ENERGY COSMIC RAYS\footnote{\sl Submitted to
Astroparticle Physics}}

\vspace{.3in}

G.~Sigl$^{a,b}$, D.~N.~Schramm$^{a,b}$ and P.~Bhattacharjee$^c$\\

\vspace{0.2in}

{\it $^a$Department of Astronomy \& Astrophysics\\
Enrico Fermi Institute, The University of Chicago, Chicago, IL~~60637-1433}\\

\vspace{0.1in}

{\it $^b$NASA/Fermilab Astrophysics Center\\
Fermi National Accelerator Laboratory, Batavia, IL~~60510-0500}\\

\vspace{0.1in}

{\it $^c$Indian Institute of Astrophysics\\
Sarjapur Road, Koramangala, Bangalore 560 034, INDIA}\\

\end{center}

\vspace{0.3in}

\begin{abstract}
In this paper we show that the conventional diffusive shock
acceleration mechanism for cosmic rays associated with
relativistic astrophysical shocks in active galactic nuclei
(AGNs) has severe difficulties to explain the highest energy
cosmic ray events. We show that protons above around
$2\times10^{20}\eV$ could have
marginally been produced by this mechanism in an AGN or a rich
galaxy cluster not further away than around $100\Mpc$. However, for
the highest energy Fly's Eye and Yakutsk events this
is inconsistent with the observed arrival directions. Galactic
and intergalactic magnetic fields appear unable to alter the
direction of such energetic particles by more than a few degrees.
We also discuss some other options for these events associated
with relativistic particles including pulsar
acceleration of high $Z$ nuclei. At the
present stage of knowledge the concept of topological defects
left over from the early universe as the source for such events
appears to be a promising option. Such sources are discussed and
possible tests of this hypothesis are proposed.
\end{abstract}
\newpage
\pagestyle{plain}
\setcounter{page}{1}
\section{Introduction}
It is generally believed that cosmic rays with energies up to
the ``ankle'' at around $3\times 10^{18}\eV$ are predominantly of
galactic origin~\cite{Gaisser93} and that energies up to around
$10^{14}\eV$ can be achieved by first order Fermi acceleration in
shocks produced by supernovae exploding into the interstellar
medium~\cite{Gaisser}. Recently the Fly's Eye detector revealed
a change in the cosmic ray composition which is correlated with
a dip in the total energy spectrum~\cite{Gaisser93} located
at the ankle.  Around this
dip the spectrum first steepens and then flattens again to a
spectral index of around $2.7$ which is even smaller than the
index of $3$ corresponding to the spectrum below the steepening.
The data are consistent
with a superposition of a steep power law spectrum of heavy
nuclei and a flatter spectrum of protons which overtakes the
former component at energies above the ankle. It is expected
that this latter high energy proton component is of extragalactic
origin. Furthermore, on 15 October 1991, the Fly's Eye observed an
event at $(3.2\pm0.6)\times10^{20}\eV$ ($1\sigma$ errors)~\cite{PS},
which is the event with the highest energy ever recorded. Interesting
enough, the world's second highest energy air shower of
$(1.1\pm0.4)\times10^{20}\eV$ was recorded at Yakutsk~\cite{Efimov,Egorov}
located within 7.8 degrees from the Fly's Eye event (see Fig.~2).
In this paper
we will assume that these events were caused by relativistic particles.

There is plenty observational evidence that AGNs and
radiogalaxies contain relativistic termination shocks which
are likely to produce high energy cosmic rays.
It therefore seems natural to extend the standard
diffusive shock acceleration scenario which works well for
supernova shocks at lower cosmic ray energies to larger
extragalactic shock scales in order to explain the origin of
this higher energetic extragalactic component. However, it turns out
to be hard to explain the highest energy events by this mechanism.
An interesting option is provided by decaying or annihilating topological
defects which could be left over from phase
transitions in the early universe at temperatures corresponding to some
Grand Unification scale. The rest of the paper is organized as
follows: We first reconsider in
section 2 the source spectrum cutoff energy $E_c$ for shock accelerated
cosmic rays as a function of shock size and average magnetic field
strenght on this lenght scale and evaluate it for some typical
observational numbers. In section 3 we discuss propagation
effects on protons and show that at least the $3.2\times10^{20}\eV$ event
is difficult to reconcile with the observational knowledge of
typical extragalactic shock parameters in this acceleration scenario.
We therefore discuss in section 4 some other options for these highest energy
events. In section 5 we suggest that such events could
alternatively be produced by topological defects. Finally we summarize
our findings in section 6.

\section{The Source Spectrum Cutoff for Extragalactic\newline
Shock Acceleration}
In relativistic shocks the cutoff energy $E_c$ for the source spectrum
of accelerated cosmic rays is in the test particle approximation
always given by $ZeBR$, the product of the charge $Ze$ of the
cosmic ray particle, the magnetic field $B$ and the size $R$ of
the shock, multiplied by some factor of order
unity~\cite{Cesarsky,Quenby} (we use natural units, {\it i.e.}
$c=h/2\pi=1$, throughout this paper). However, it turns out that for the
highest energies the mean free path of the particle becomes
comparable to the shock size $R$, which sometimes is not
properly accounted for. We therefore calculate here our own
approximation for $E_c$.

\subsection{The Source Energy Cutoff}
The acceleration of a cosmic ray particle of energy $E$ in an
astrophysical shock is governed by the equation
\begin{equation}
  {dE\over dt}={E\over T_{acc}}\,,\label{Eacc}
\end{equation}
where $T_{acc}$ is the energy dependent acceleration time. In
the statistical point of view the slope $q$ of the energy
spectrum $dN/dE\propto E^{-q}$ of the particle flux is related
to $T_{acc}$ and $T_{esc}$, the mean (in general also energy
dependent) escape time by~\cite{Gaisser,Cesarsky}
\begin{equation}
  q=1+{T_{acc}\over T_{esc}}\,.\label{q}
\end{equation}
For first order Fermi acceleration at nonrelativistic shocks
caused by supernovae, $T_{acc}$ is usually given by
\begin{equation}
  T_{acc}={3\over u_1-u_2}\left({D_1\over u_1}+
          {D_2\over u_2}\right)\,,\label{Tacc}
\end{equation}
where $u_1$, $u_2$ are the up- and downstream velocities of the
shock and $D_1$ and $D_2$ are the corresponding diffusion
coefficients, respectively. Diffusion is dominated by magnetic
pitch angle scattering caused by inhomogeneities in the magnetic
field~\cite{Blandford,Cesarsky2}. Therefore, the mean free path $\lambda$
is bounded from below by some multiple $g$ of the gyroradius
$r_L=E/(ZeB)$ and $D_1$ and $D_2$ can for ultrarelativistic
particles be estimated by
\begin{equation}
  D_1,D_2\sim\lambda/3\ga{gE\over 3ZeB}\,.\label{Diff}
\end{equation}
For nonrelativistic shocks, $g$ is usually set equal to
1~\cite{Gaisser,Cesarsky}. However, as we deal with the highest
energetic extragalactic cosmic ray component, we have to
consider relativistic shocks because they provide the most powerful
accelerators. Monte-Carlo simulations of such
relativistic shocks yield $g\sim40$~\cite{Quenby,Lieu}.
Furthermore, the acceleration turns out to be enhanced compared
to eq.~(\ref{Tacc}) by about a factor of 10 in highly
inclined~\cite{Lieu} and by about a factor of 13.5 in
parallel~\cite{Quenby89} relativistic shocks, respectively.

Putting everything together and maximizing $T_{acc}$ from
eq.~(\ref{Tacc}) we arrive at
\begin{equation}
  T_{acc}\ga{g\over2.25}{E\over ZeB}\,,\label{Taccmin}
\end{equation}
On the other hand, as long as the diffusion approximation is
valid, {\it i.e.} as long as $\lambda<R$ corresponding to
$E<E_{diff}\equiv ZeBR/b$, the escape time is given by
$T_{esc}=R^2/\lambda$, whereas for $E\geq E_{diff}$ the particles
are freely streaming out of the shock region to a good approximation
and $T_{esc}=R$. Using eqs.~(\ref{q}) and (\ref{Taccmin}), we
thus get
\begin{equation}
  q(E> E_{diff})\sim1+{E\over2.25E_{diff}}\,.\label{gam}
\end{equation}
Defining the cutoff energy $E_c$ as the energy where the source
spectral index becomes 3 (remember that the slope of the energy
spectrum observed at the earth was around 2.7 in the region of
highest energies) yields
\begin{equation}
  E_c\equiv E_{q=3}\sim10^{17}\eV\,Z
     \left({R\over{\rm kpc}}\right)
     \left({B\over\mu{\rm G}}\right)\,.\label{Ec}
\end{equation}
This is compatible or even higher as compared to similar
estimates~\cite{Cesarsky,Quenby,Hillas,Biermann,Rachen}. We have
assumed here that the magnetic field is parallel to the shock
normal. If that is not the case there will be an electric field
${\bf E}={\bf u}\times{\bf B}$ in the shock rest frame (${\bf
u}$ is the shock velocity in the lab frame). This causes drift
acceleration of charged particles to a maximal energy given by
\begin{equation}
  E_{max}=ZeuBR\sim10^{18}\eV\,Zu
  \left({R\over{\rm kpc}}\right)
  \left({B\over\mu{\rm G}}\right)\,,\label{Emax}
\end{equation}
which is around one order of magnitude larger than
eq.~(\ref{Ec}) if $u$ approaches the speed of light.
However, the electric field ${\bf E}$ is expected
to be much smaller in general due to plasma effects so that
rather special conditions have to be fulfilled in order that
such high energies can be approached.

Throughout the rest of this section and the next section we will
restrict our discussion to protons ($Z=1$). We will comment on
nuclei as possible candidates for events with energies above
$10^{20}\eV$ in section 4.3.

\subsection{The Cutoff in Numbers}
Let us now look at some observational numbers for $R$ and $B$ and
evaluate the corresponding cutoff energy for a proton.
Cesarsky~\cite{Cesarsky} cited the example of a galaxy
encounter (NGC 4038/39~\cite{ob1}) as a location of a strong relativistic
shock with a magnetic field of about $40\muG$ on a scale of
about $2\kpc$, leading to $E_c\sim8\times10^{18}\eV$, significantly too small
to explain the origin of cosmic rays with energies as high as
the Fly's Eye event of $3\times10^{20}\eV$. Potentially more
interesting candidates for
acceleration beyond $10^{20}\eV$ are revealed by the ``hot
spots'' and radio lobes of CygA
with $B\sim400\muG$, $R\sim3\kpc$ and $B\sim4\muG$,
$R\sim300\kpc$, respectively~\cite{ob2}, cited by
Quenby~\cite{Quenby} which lead to a cutoff energy
$E_c\la5\times10^{20}\eV$. There are also some indirect
indications from gamma ray astronomy that in some quasars protons
could be shock accelerated to energies of about
$2\times10^{20}\eV$~\cite{Mannheim}.

There is one nearby object where pure application of
formula~(\ref{Ec}) on observational indications for $R$ and $B$ leads to
an $E_c$ larger than
$3\times10^{20}\eV$. This is if one takes the whole Virgo cluster
with an extension of $R\sim10\Mpc$ and a intracluster magnetic
field $B\sim1.5\muG$ which is compatible with
observations~\cite{Cesarsky,ob3} and leads to
$E_c\sim1.5\times10^{21}\eV$. However, it is highly improbable that
the whole Virgo cluster moving through intercluster space forms
a relativistic shock of such an enormous extension, effective in
coherent cosmic ray particle acceleration.

Based on examples of the sort presented above there is some
common belief that for protons the highest source energy
achievable by diffusive shock acceleration in quasars and
radiogalaxies is around
$10^{21}\eV$~\cite{Biermann,Rachen,Wolfendale}. We will nevertheless
show in the next section that even if this is true, it
is still difficult to explain the observed Fly's Eye event because of
the information we have on its arrival direction.

\section{Propagation Effects on Protons above 100 EeV}
\subsection{Overview}
Up to now we were only talking about the source energy spectrum.
However, a proton traveling through space is in general subject
to interactions, mainly with photons and magnetic fields. The
latter effect leads to a curved path with a radius given by the
gyroradius
\begin{equation}
  r_L=1.1{\rm pc}\left({E\over10^{15}\eV}\right)
  \left({B\over\mu{\rm G}}\right)^{-1}\,.\label{gyro}
\end{equation}
The former effect leads to scattering and an effective energy
loss as long as the proton energy lies above a kinematical
threshold energy
\begin{equation}
  E_{th}={W_{th}^2-m^2_N\over2\epsilon(1+\cos\theta)}\,,\label{Eth}
\end{equation}
depending on the angle $\theta$ between the incoming photon
momentum and the negative nucleon momentum.
Here $m_N$ is the nucleon mass, $\epsilon$ is the typical energy
of an incoming photon and
$W_{th}$ is the center of mass energy threshold for the
particular reaction under consideration. The most important ones
are electron pair production $N+\gamma\to N+e^++e^-$ with
$W^{e^+e^-}_{th}=m_N+2m_e$, and pion production, $N+\gamma\to
N+\pi$, with $W^\pi_{th}=m_N+m_\pi$, with $m_e$ and
$m_\pi$ the electron and pion mass, respectively. For a fraction
of its propagation time depending on the effective neutron
lifetime the proton will actually transform into a neutron which
does, however, not have much influence on this treatment as
protons and neutrons have similar interactions with the MBR. Thus,
nucleons interact with photons of the microwave background
radiation (MBR) producing $e^+e^-$ pairs and pions above an energy of
about $5\times10^{17}\eV$ and $1.1\times10^{20}\eV$, respectively.
The latter effect is known as the Greisen-Zatsepin-Kuz'min (GZK)
effect~\cite{Greisen}. There are also plenty of infrared and
optical photons around luminous AGNs and galaxy clusters and
especially in their central regions leading to
correspondingly smaller nucleon threshold energies. However,
their number density is in general not much bigger than that
of the MBR photons and the cross section at the correspondingly
higher center of mass energy is even smaller. This implies that
possible interactions with these higher energy photons does
not substantially reduce the mean free path
below $6\Mpc$, the typical mean free
path in the MBR~\cite{Cronin}. As this is much larger than
typical galaxy sizes, we will restrict our considerations to
interactions with the MBR. That means that we are on the safe
side and get lower limits on energy losses and upper limits
on the corresponding possible travel distances~\cite{Pollak}.

\subsection{Energy Losses}
We now ask the question in what distance range a source causing
nucleon induced
events of energy $E_0$ on earth could be if the maximal source
energy is $E_s$. To this end we introduce the ``longitudinal
nucleon energy'' ${\tilde E}_N=(1+\cos\theta)E_N/2\ga E_{th}$
besides the nucleon energy $E_N$ in the comoving frame. The last
inequality holds because of eq.~(\ref{Eth}) as long as reactions
are kinematically allowed. By performing a Lorentz
transformation from the
comoving frame to the center of mass frame corresponding to a
gamma factor $\gamma_{cm}\sim\tilde{E}_N/W=\tilde{E}_N/(m^2_N+
4\epsilon\tilde{E}_N)^{1/2}$ one can see that after scattering
the nucleon energy in the comoving frame is given by
\begin{equation}
  E^\prime_N\sim\gamma_{cm}(E^\prime_{cm}+p^\prime_{cm}
  \cos\theta_{cm})\sim{\tilde{E}_N\over W}
  (E^\prime_{cm}+p^\prime_{cm}\cos\theta_{cm})\,,\label{Eprime}
\end{equation}
where $p^\prime_{cm}$ and $E^\prime_{cm}$ are momentum and energy
of the nucleon after scattering and $\theta_{cm}$ is the
scattering angle, evaluated in the center of mass frame,
respectively. Above $10^{20}\eV$ the energy loss is dominated by
pion production~\cite{Puget} for which
these quantities are related to $W$ by $(m_N^2+
p^{\prime 2}_{cm})^{1/2}+(m_\pi^2+p^{\prime 2}_{cm})^{1/2}=W$. As
we are only interested in an estimate we set $\tilde{E}_N=E_N/2$
and neglect the energy dependence of the mean free path (which
in the energy range we are interested in is a good
approximation~\cite{Cronin}) $\lambda\sim6\Mpc$ in the following
calculation. Because of eq.~(\ref{Eprime}) the energy change in
a scattering event $\xi(E,\theta_{cm})$ relative to the lower energy
is given by
\begin{equation}
  \xi(E,\theta_{cm})={W\over E^\prime_{cm}+p^\prime_{cm}
  \cos\theta_{cm}}-1\,.\label{xi}
\end{equation}
Furthermore, we neglect energy loss due to cosmological redshift
as we are dealing with non-cosmological distances at these high
energies.
By integrating from lower to higher energies one can show that
the mean $\bar l$ and the variance $\Delta l^2$ of the distance
as a function of $E_0$ and $E_s$ can be estimated by
\begin{equation}
  \bar l(E_0,E_s)\sim\lambda\int_{E_0}^{E_s}{dE\over E\bar\xi(E)}\,,
  \quad\Delta l^2(E_0,E_s)\sim\lambda\bar l(E_0,E_s)+
  {\lambda^2\over 3}\int_{E_0}^{E_s}{dE\over E\bar\xi^3(E)}
  \left({p^\prime_{cm}\over E^\prime_{cm}}\right)^2
  \,.\label{esti}
\end{equation}
The first term in the variance is due to the fluctuation of the
number of scatterings and the second one is due to the
fluctuation of the energy transfer $\xi$ of eq.~(\ref{xi})
around its mean $\bar\xi(E)\sim W/E^\prime_{cm}-1$ averaged over
the center of mass scattering
angle $\theta_{cm}$. We have numerically integrated
eq.~(\ref{esti}). The results are shown in  Fig.~1a and 1b.
We see that for $E_0=1.7\times 10^{20}\eV$, the lowest possible
energy for the Fly's Eye event the distance must be
smaller than $\sim100\Mpc$ and $\sim130\Mpc$ on the $3\sigma$
level for a source energy $E_s=10^{21}\eV$ and $E_s=10^{22}\eV$,
respectively. For the best fit energy of $3.2\times10^{20}\eV$ we
get $\sim60\Mpc$ and $\sim90\Mpc$ for the corresponding $3\sigma$
upper limits of distance for the same source energies.

Thus from the energy point of view an AGN or a galaxy cluster constituting
a large scale shock with the intercluster medium could be marginally
able to cause events like the $3\times10^{20}\eV$ Fly's Eye event by shock
accelerated protons, if it is not much further away than $100\Mpc$.
Note that this number actually means the path length for which
the distance is a lower limit which can be overtaken if the path
is curved.
However, we now show that the arrival direction of such a nucleon
would then have to lie within about 10 degrees of the direction of
its source if conventional wisdom about magnetic fields is used.

\subsection{Deviations from Rectilinear Propagation}
Let us first discuss deflection caused by magnetic fields. Unfortunately,
not much is known about extragalactic magnetic fields. Faraday
rotation measurements of extragalactic radio sources seem to
suggest fields of the order of $10^{-9}\G$ which could be homogeneous
on large scales~\cite{Parker}. Most estimates are of this order
or below~\cite{cosB}, a more recent one being as low as
$3\times10^{-11}\G$~\cite{Asseo}. The bending angle
$\alpha$ in radian for a proton traveling in a magnetic field
satisfies $\alpha\la\int dl/r_L$
where $dl$ is the differential path length and $r_L$ is given by
eq.~(\ref{gyro}). A $10^{-9}\G$ field leads to a maximal
bending angle of around 10 degrees for a proton with arrival energy of
$3\times10^{20}\eV$. This maximum can only be reached if the
magnetic field is perpendicular to the proton path and does not
change its polarization considerably on a scale of $\sim100\Mpc$.
Otherwise the bending angle is reduced at least by a factor
$(d_c/100{\rm Mpc})^{1/2}$ where $d_c$ is the magnetic field
coherence length scale. Therefore, even if the typical intercluster
field would be as high as $10^{-8}\G$~\cite{Vallee} the bending
angle would still not be larger than $10^o$ if the coherence
length scale is $\sim1\Mpc$.

A proton can also be deflected
by our own galactic magnetic field which is of the order
of $3\muG$~\cite{Parker,Vallee2}. Based on radio telescope
observations of Faraday rotations~\cite{Vallee2} its coherent
component is supposed
to have a cylindrical structure of diameter $\sim25\kpc$
and height of order $1\kpc$ and being polarized in the
direction of decreasing galactic longitude in the outer region.
The random component is supposed to be of the same order of
magnitude. Given the arrival direction
of the Fly's Eye and the Yakutsk events shown in
Fig.~2 we see that the corresponding
path length through the field is less than $3\kpc$ resulting in a
maximum bending angle of $\sim2^o$.

We finally remark that even though magnetic fields in galaxy
clusters as high as $10^{-6}\G$ as already
mentioned~\cite{ob3} could cause a significant
deflection this does not influence our argumentation. It
only means that if our proton encountered such ``magnetic lenses''
it should point approximately to the last
encountered. There still has to be
a nearby galaxy cluster in its arrival direction. Furthermore,
the above mentioned fact that a proton partly transforms into a
neutron during propagation even tends to decrease the bending further.

The propagation direction could in principle also be
changed by scattering with photons, but the scattering angles
involved are much too small. To see that we use our notation
from above and note that the scattering
angle $\theta_{sc}$ in the comoving frame obeys
\begin{equation}
  \tan\theta_{sc}\la{p^\prime_{cm}\sin\theta_{cm}\over
  \gamma_{cm}(p^\prime_{cm}\cos\theta_{cm}+E^\prime_{cm})}\,.\label{sc1}
\end{equation}
Maximizing with respect to $\theta_{cm}$ and using
$p^\prime_{cm}\leq(W^2-m^2_N)^{1/2}\leq(4\epsilon\tilde{E}_N)^{1/2}$
leads to an estimate independent from the final state $X$ in the
reaction $N+\gamma\to N+X$,
\begin{equation}
  \tan\theta_{sc}\la\left[{4\epsilon\over\tilde{E}_N}
  \left(1+{4\epsilon\tilde{E}_N\over m^2_N}\right)\right]^{1/2}
  \la10^{-11}\,,\label{sc2}
\end{equation}
where we have used $\epsilon\sim3\times10^{-4}\eV$ and
$\tilde{E}_N\ga3\times10^{20}\eV$. Because the mean free path at
these energies is of order $6\Mpc$~\cite{Cronin} we expect
only a few scattering events during traveling over a distance of
$100\Mpc$ which thus never can lead to a significant change in the
propagation direction.

The arrival direction of the Fly's Eye event is given by
$\alpha=85.2^o\pm1^o$ and $\delta=48^o\pm10^o$. No potentially
interesting object with a sufficiently powerful shock
acceleration engine of the scale discussed in the
previous section is located within $100\Mpc$ in that direction.
This can be seen from Fig.~2 where we show the directions
to important nearby galactic objects, galaxy clusters and AGNs.
One of the three prominent FR-II radio galaxies listed in~\cite{Rachen}
and thought to contribute to the proton spectrum between $10^{18}\eV$
and $10^{20}\eV$ ({\it i.e.} below the GZK threshold), namely
3C111, has the coordinates $\alpha=63.75^o$, $\delta=37.9^o$ and lies thus
$(18.6\pm5.1)^o$ away from that direction. But it is at least
$140\Mpc$ away. The only quasars within around $10^o$ of the arrival
direction are 3C147 and 3C159~\cite{catalog}, both of them at a distance
of at least $900\Mpc$. The only Seyfert galaxy within $10^o$ is MCG 08.11
at an angular distance of $(2.35\pm8.4)^o$ and a distance of at least
$60\Mpc$~\cite{catalog}. Although this is much nearer than the before mentioned
quasars it produces a much lower radio flux at the earth and therefore
seems also not likely to produce high cosmic ray fluxes.
Under the quite improbable assumptions discussed at the end of the previous
section the Virgo cluster seems to be the nearest possible candidate.
However, it is located around $90^o$ away from that direction.

\section{Other Options for Ultrahigh Energy Events}
We saw in the previous section that especially the Fly's Eye
event is difficult to explain as a proton within the standard
shock acceleration scenario. We therefore now like to discuss some
other options beginning with secondary particles produced by
shock accelerated protons.

\subsection{Secondary Photons}
Is the shower development of the highest energy Fly's Eye event
consistent with what would be expected if it was caused by a
photon? At these high energies the photon begins to interact
with the earth's magnetic field already {\it above} the atmosphere.
Fitting the shower shape with a three parameter Gaisser-Hillas
shower development function~\cite{GH} gives a depth of first interaction
of $-100\gcm2$ and thus seems to indicate a first interaction
above the atmosphere. However, it could also indicate that
the fitting function is simply inappropriate at these high energies.
Indeed, it is now believed that fitting heavy nuclei induced
showers can lead to similar negative values~\cite{PS}.
Taking the LPM effect into account the average shower maximum
is expected to be somewhat larger than for a proton induced
shower with an
average width about twice as large as the corresponding proton
profile width~\cite{Proc}. As the fluctuations are expected to be
large this could still be compatible with the reported maximum
at $(815\pm50)\gcm2$.

Photons of such a high energy have a secondary
origin in the standard scenario as they are produced by decay of
pions or $e^+e^-$ interactions which in turn are produced by the
interactions of the
cosmic ray protons with the MBR. The photon mean free path becomes
comparable with or larger than that of the protons above a few
$10^{19}\eV$~\cite{Wdowczyk}. The exact value of the photon
to proton ratio depends on the universal radio background
and the intergalactic magnetic field strength. The former leads to
additional losses due to electron pair production. The latter leads
to an inefficient electromagnetic cascade development $\gamma+\gamma_b
\to e^+e^-$, $e^+e^-+\gamma_b\to e^+e^-+\gamma$ ($\gamma_b$ is the
background photon) due to synchrotron cooling of the electrons
even for fields as low as $10^{-10}\G$. Typical estimates for
the photon to proton ratios are considerably smaller than 1 above
$10^{20}\eV$~\cite{Stecker,Ah1}. That makes the problem even harder as
there have to be (even more abundant and of higher energy)
primary protons acting as the source for such photons.
Wolfendale~\cite{Wolfendale} claims that the photon to proton ratio
could exceed unity above $\sim3\times10^{20}\eV$ if the source
energy cutoff $E_c$ is much above $10^{21}\eV$. This possibility
thus runs into trouble with our discussion of $E_c$ in section 2.

\subsection{Secondary Neutrinos}
Could the highest energy Fly's Eye event have been an extragalactic
neutrino produced as a secondary of a shock accelerated proton?
Because neutrinos essentially lose no energy apart from redshift in
going over cosmological distances it could have been produced
by a proton interacting near its acceleration site thus avoiding excessive
subsequent energy losses due to downscattering.
However, it turns out that the neutrino yield above $10^{18}\eV$
is considerably smaller than one for all reasonable injection spectra.
Furthermore, at the highest energies the spectral index observed at
the earth is predicted to be 0.5 larger than the corresponding
proton spectral index~\cite{Schramm}. As
even at these energies the neutrino nucleon cross section is
still by a factor of at least $10^6$ smaller than the nucleon nucleon
(and also the gamma nucleon) cross section the average interaction
depth is much larger than the atmospheric depth. All that leads
to the conclusion that the event rate due to neutrinos should
be much smaller than that due to protons at the same energy.
One would also expect the neutrino induced showers to start
predominantly at high depths {\it i.e.} near the horizon~\cite{Stecker}.
Therefore, if a neutrino caused the event then it was a very atypical
one and we would expect much more proton and even photon
events at the same energy.

\subsection{Secondary Neutrons and Heavy Nuclei}
In the shock acceleration scenario neutrons are also produced as
secondaries of protons or heavy nuclei as they are neutral and can not
be accelerated directly. Furthermore, due to instability they have
only a finite range which for $3\times10^{20}\eV$ is about $3\Mpc$.
The puzzle is therefore not solved if our events are caused by neutrons.

Looking at eqs.~(\ref{Ec}) and (\ref{Emax}) one realizes that heavy nuclei can
reach maximal energies which are higher by a factor $Z$ compared
to the protons. However, heavy nuclei lose energy not only due
to the processes which dominate the energy loss of protons but
also due to the giant dipole resonance which leads to
photodisintegration. Above $10^{20}\eV$ the corresponding energy
loss rates are about a factor 10 higher than those for
protons~\cite{Puget} and are typically due to proton stripping
reactions. For example, a $^{56}$Fe nucleus
being launched with an energy of $10^{21}\eV$ will be below
$10^{20}\eV$ after traveling $8\Mpc$~\cite{Cronin}. Thus, for our
purposes nuclei
are only interesting when they are of galactic origin.
There are two galactic sites which could provide acceleration
to interesting energies for heavy nuclei. The first one is the
termination shock of the galactic wind caused by the milky
way~\cite{Morfill}. This leads to maximal energies of $\sim
3\times10^{17}\eV\,Z(u/600\km\sec^{-1})^2(T_A/1.5\times10^{10}{\rm y})
(B/0.1\muG)$~\cite{Gaisser} where $u_w$ is the galactic wind speed and
$T_A$ is the shock lifetime. Even for $Z\sim100$ this is
significantly too low and one is forced to use quite extreme
parameters to reach beyond $10^{20}\eV$. The second site would be
even more natural to produce predominantly high energy heavy
nuclei, namely young supernova remnants which form a pulsar wind
shock~\cite{Gaisser}. This is because pulsars can have quite
large surface magnetic fields of order $10^{12}\G$ leading to
fields of order $10\G$ on scales of $10^{14}$cm~\cite{Gaisser}. The
pulsar wind can be relativistic so that application of
eq.~(\ref{Emax}) leads to $E_{max}\sim10^{18}\eV\,Z$ which for
$Z\sim50$ is about one order of magnitude smaller than
the highest observed energies. Adopting the more conservative
estimate eq.~(\ref{Ec}) leads to a short fall of about two orders
of magnitude.

Indeed, as can be
seen from Fig.~2, the Crab nebula lies near the arrival direction
of our events
and is expected to have magnetic fields of the order of
$10^{-3}\G$ on a scale of $2\pc$~\cite{Chevalier} leading to
estimates for the energy cutoff quite similar to the above
mentioned. However, a calculation
including the uncertainty of the latter one gives a relative angle
of $(25.9\pm7.5)^o$ which is more than $3\sigma$ away.
Furthermore, taking into account
deflection effects due to the coherent galactic magnetic field component
mentioned in section 3.3 increases the angular distance as
the path should be bent towards the galactic north pole (see Fig.~2).
For example,
for $Z=5$ the angular difference from the arrival direction would be
$(29.5\pm6.9)^o$. It thus seems that only large
bending by almost $360^o$ could explain a possible origin from the
Crab. For $Z\ga50$ the Larmor radius is
$\la2\kpc$ at these energies which indeed comes near the
required amount of bending.
In that case, however, the arrival direction is not
expected to be correlated with the source location in a simple
way and the propagation of these particles should better be considered
as diffusion in the magnetic field. The source could therefore be
any galactic site being able to produce the required source energy and
this possibility can not be completely excluded although, as mentioned
above, current models fall short in energy. It should be noted
that the shower maximum at $(815\pm50)\gcm2$ allows no definite
distinction between a proton and a heavy nucleus induced shower
as the expected numbers for these options are $\sim850\gcm2$ and
$\sim775\gcm2$ for iron, respectively~\cite{Gaisser,PS}.

There are still problems left in interpreting the $3\times10^{20}\eV$
event as caused by a heavy nucleus.
As so it is possible that heavy nuclei could be disintegrated
already at the source of acceleration~\cite{Pollak}. Furthermore, the
Fly's Eye data between $10^{18}\eV$ and $10^{20}\eV$ suggest
the transition to a lighter component as we already mentioned.

\subsection{Some Other Options}
There were some other suggestions how one could get to higher
source energies. For example, Colgate~\cite{Colgate} claimed that in
the relativistic plasma of AGN jets energies as high as
$10^{24}\eV$ could be reached due to a plasma pinch effect
similar than that used in Tokamaks. Note however that due to
Fig.~1a and 1b the source of the highest energy events could still not
be much further away than $150\Mpc$. Because the Larmor radius
grows with energy the possible bending angle caused by magnetic
fields could also not be enhanced significantly beyond 15
degrees so the problem remains.

It was suggested~\cite{Hayakawa} that high energy events could
be caused by relativistic dust grains. The lateral shower
profile caused by a dust grain entering the atmosphere is expected
not to show a broad maximum but instead to have a more or less
constant lateral spread as long as the grain remains large and energetic
enough to produce secondary particles. However,
the highest energy Fly's Eye event showed a quite ``normal'' shower
development typical for a primary proton or possibly a photon
which leaves the dust grain hypothesis to seem not very likely.
Furthermore there is a tendency for
these grains to break up by interactions with photons and gas
atoms in the interstellar medium~\cite{Berezinsky}.

\section{Cosmic Rays Produced by Topological Defects}
Topological defects (TDs)~\cite{Vilenkin} could have been formed in the early
universe during phase transitions associated with spontaneous
breaking of symmetries implemented in unified models of high energy
interactions. Such TDs are magnetic monopoles, cosmic strings, domain
walls, superconducting cosmic strings, textures, etc. TDs are
topologically stable but can nevertheless be destroyed due to
physical processes like collapse or annihilation~\cite{Hill1,Hill2,Bh1,Bh2}.
In that
case the energy stored in the defects is released in the form of
massive quanta of the fields like gauge fields and Higgs fields
associated with the broken symmetry. These ``X'' particles released
from the TDs would typically decay into quarks and leptons.
Hadronization of the quarks would produce jets of hadrons containing
mainly light mesons (pions) together with a small fraction
($\sim3\%$) of nucleons. The gamma rays and neutrinos from the decay
of the pions would thus be the dominant particles in the final decay
products of the X particles. The mass $m_X$ of the X particles is
typically of the order of the symmetry-breaking scale which in Grand
Unified Theories (GUTs) can be $\sim10^{25}\eV$, or even the Planck
scale $\sim10^{28}\eV$. The decay of the X particles released from TDs
can thus give rise to nucleons, gamma rays and neutrinos with
energies up to $\sim m_X$, very much higher than what can be achieved
by astrophysical shock acceleration mechanism. The cosmic ray
particles can thus be produced directly in this scenario, and no
acceleration mechanism is needed.

The production spectra of the nucleons, gamma rays and neutrinos in
the TD scenario are determined by the physics of fragmentation of
quarks into hadrons. Extrapolation~\cite{Hill1} of QCD based
hadronization models (which describe well the GeV scale collider
data) to the extremely high energies gives a power-law
approximation~\cite{Bh1,Bh2,Bh3}
to the differential production spectra with a power-law index
$q\sim$1.32 for nucleons as well as pions. The decay of the neutral pions
thus gives a differential gamma ray production spectrum also with
$q$=1.32. It is to be emphasized, however, that there
is a great deal of uncertainty in extrapolating the low energy QCD
models of hadronization to the extremely high energies involved in
the present situation. Moreover, the gamma ray production spectrum
can be somewhat different from the proton production spectrum if one
considers the gamma rays generated by the charged leptons (electrons
and positrons) in the primary decay products of the X particles.
The electrons and positrons coming from the decay of the
charged pions in the hadronic jets also contribute to the overall
primary gamma rays. The main point, however, is that the
production spectra of cosmic ray particles in the TD scenario can in
principle be considerably flatter than in the standard shock
acceleration scenario. The latter, to recall, by and large produces
differential production spectra with $q>2$.

One consequence of a relatively flat production spectrum in the TD
scenario would be the ``recovery''~\cite{Bh3} of the evolved proton spectrum
after the GZK ``cutoff''~\cite{Greisen}. While this is heartening
from the point
of view of prospects for detecting protons above the GZK ``cutoff'',
too flat a proton spectrum may cause problems in that it may give
rise to excessive gamma ray flux at much lower energies, as discussed
below. In any case, as first discussed in Ref.~\cite{Ah1}, the
photon-to-proton ratio in the evolved spectra can be considerably
larger than 1 above $10^{20}\eV$ in the TD scenario~\cite{Ah1} (because of
the {\it primary}
gamma rays which outnumber the protons by a factor of at least 10 at
production, and also because of higher transparency of gamma rays
relative to the protons at these energies), and so the cosmic rays
above $10^{20}\eV$ are predicted
to be mainly {\it primary} gamma rays rather than protons.

Gamma rays as well as protons of ultrahigh energies
generate lower energy gamma rays by $\gamma- \gamma_b$ and
$p-\gamma_b$ collisions with the photons ($\gamma_b$) of the
background radiation fields. The electromagnetic component of the
energy lost by the photons and protons in these collisions cascades
down to lower energies by electromagnetic cascading in the universal
radio background (URB), the microwave background (CMBR), and in the
infrared background (IRB) (in order of decreasing energy of the
propagating photon). Recently it has been realised~\cite{Chi1,Chi2,Ah2}
that the measured flux of extragalactic gamma rays in the $100\MeV$
region~\cite{Fichtel}
provides an upper limit on the total energy density of the
cascade-initiating
electromagnetic radiation that can possibly be released in the
universe due to $p-\gamma_b$ and $\gamma-\gamma_b$ interactions.
This, in turn, restricts the shapes of the proton as well as the
primary photon spectra in the highest energy region.
The authors of Ref.~\cite{Chi1} claim that
a proton spectrum with $q=1.32$ at injection and extending to
$10^{24}\eV$ would by itself give rise to a $100\MeV$ gamma ray flux
exceeding the measured flux by a factor of 2 if the evolved proton
spectrum is normalized~\cite{Ah1,Bh3} to observed particle flux at
$4\times10^{19}\eV$.
This question has recently been studied in detail~\cite{Ah2} by a careful
numerical calculation of the cascading process including the gamma
rays generated by
both $p-\gamma_b$ as well as the $\gamma-\gamma_b$ processes.
It is found that whether or not the predicted $100\MeV$ gamma ray flux
exceeds the measured value depends strongly on the level of the IRB
as well as on its cosmological evolutionary history both of which are
rather uncertain, and so a firm conclusion in this regard cannot be
drawn at this stage. Nevertheless, the authors of Ref.~\cite{Chi2} have
suggested that the possible problem arising from requirement of
consistency with the measured $100\MeV$ gamma rays can be avoided if
the cosmic rays above $10^{19}\eV$
are mainly gamma rays and not protons. The preliminary analysis of
Ref.~\cite{Chi2} shows that this is possible provided the injection spectrum
of the primary gamma rays above $10^{20}\eV$ in the TD model is made
somewhat steeper ($q\sim2.4$) compared to the protons ($q\sim1.5$) and
$\gamma/p$ is demanded to be $\sim60$ at injection so that the proton
component is made
negligible compared to the photons. The average multiplicity in the
hadronic jets arising out of the decay of the X particles is also
required to be somewhat higher than what naive extrapolation of the
low energy QCD based models of jet fragmentation indicates. While all
these phenomenological requirements need to be substantiated on more
theoretical grounds, the general conclusion that seems to arise from
the above discussion is that the highest energy cosmic ray particles
in the TD scenario {\it should be} mainly gamma rays and not
protons. And, of course, primary neutrinos~\cite{Bh3} should be at least as
abundant as the gamma rays, perhaps even more.

Could the $3\times10^{20}\eV$ Fly's Eye event be a primary gamma ray
due to TD collapse or annihilation? As already mentioned in
Section 4.1 above, the shower development is not in contradiction
with what is expected for a primary gamma.
The resulting electromagnetic shower can
in fact be very similar~\cite{Ah3} to a proton-induced shower, although some
differential parameters, e.g., muon/electron ratio at large distances
from the core of the shower can in principle be used for effective
separation~\cite{Ah3} of these photon-induced showers from the proton-induced
ones.

How could one distinguish between the TD option and the galactic heavy
nuclei hypothesis which seems to be the least problematic option
within the standard picture?
Heavy nuclei are expected to produce substantially more
muons compared to gammas of the same energy~\cite{Gaisser}.
It should therefore also be possible to draw a decision between these
options as more statistics
is available at these highest energies.

The lack of any obviously identifiable
astrophysical source for the event is not a problem for the TD
scenario because TDs are not necessarily expected to be associated
with any astrophysical sources such as galaxy clusters or AGNs.
The TD model thus seems to offer an attractive option in this regard.

It is, however, expected that the same TD annihilation event would
also produce {\it lower energy} gamma rays which would arrive at
earth at roughly the same time and with same arrival direction as
the $3\times10^{20}\eV$ Fly's Eye event. Unfortunately, the CASA
array~\cite{Casa}
capable of detecting such gamma rays was not operating at the
time when the above Fly's Eye event was recorded. However, the CYGNUS
array~\cite{Cygnus} capable of detecting gamma rays above about $100\TeV$ was
operating and it detected no event~\cite{Hoffmann} that can be associated with
the Fly's Eye event. If the {\it integral} primary gamma ray spectrum
between $10^{14}\eV$ and $10^{20}\eV$ due to TD annihilation is taken
to be approximately proportional to $E^{-\alpha}$, ($\alpha>0$),
and if one (optimistically) takes the flux at $10^{20}\eV$ as $\sim$ 1
per $1000\km^2$, then above $10^{14}\eV$ and in an area of $0.02\km^2$
(roughly the area of the CYGNUS array) one would expect an integral
flux $F(E>10^{14}\eV)\sim2\times10^{6\alpha-5}$ events per $0.02\km^2$.
The non-detection by the CYGNUS array of any gamma ray event in the $100\TeV$
region coincident with the Fly's Eye event can then be
interpreted in terms $\alpha$ being $\la0.78$ (i.e., a relatively flat
spectrum) in the TD model at
energies below $10^{20}\eV$. (Note that in conventional scenarios
$\alpha$ is usually taken to be $\sim$ 1). For example, if in the TD
model one takes $\alpha\sim$ 0.32~\cite{Bh3} and neglects the attenuation
due to interaction with the CMBR (thus overestimating the expected
flux), then $F(E>10^{14}\eV)\sim0.0017$ events in $0.02\km^2$, and so
CYGNUS may have missed the event. This point,
however, needs further investigation, and will be discussed elsewhere.

\section{Conclusions}
We are lead to the conclusion that protons arriving at the earth
with energies of $3\times10^{20}\eV$ or above are very likely to have come from
an AGN or a galaxy cluster not further away than $100\Mpc$ if
they were produced there via the standard diffusive shock
acceleration mechanism. Even then the necessary conditions to be
fulfilled in such relativistic strong shocks seem highly
improbable as long as the shock parameters have to be compatible
with observational data. Furthermore, the arrival direction of
such protons have to be within around 15 degrees in the
direction of their source. The $3\times10^{20}\eV$
Fly's Eye event and the highest energy Yakutsk event were
therefore very likely not protons produced within the
standard diffusive shock acceleration scenario as they do not point
to some possible source being nearer than $100\Mpc$.

Some other explanations for such events like that being produced
by secondaries of shock accelerated protons were discussed. Within the
astrophysical shock scenario the most promising, although also
problematic option seem to be heavy nuclei of galactic origin
which could be accelerated in pulsar wind shocks driven by young
supernova remnants.

We therefore conclude that at least some improvements in the
understanding of the current acceleration picture have to be
made in order to explain the highest energy cosmic rays
observed. It seems possible that a completely new
production mechanism for such particles is necessary. We suggested
that the TD model could be a promising option. It is curious
that such an exotic option seems to have less difficulties in
explaining
these ultra-high energy particles. All other current options
appear to require suspension of belief in seemingly well
substantiated observational numbers or indicate incomplete
understanding of the underlying physical process. We list all
the options discussed here
together with their problems in Table~1.

\bigskip

\centerline{\bf Acknowledgements}

\bigskip

We would like to thank James Cronin, Eugene Parker,
Hongyue Dai, Stirling
Colgate, Gene Loh, Cy Hoffman and Corbin Covault
for highly valuable discussions. In addition, we are grateful
to Paul Sommers and Rene Ong for their suggestions concerning
the manuscript.
We would also like to thank Chris Hill for his collaboration in
earlier papers dealing with topological defects and his enthusiastic
contributions to this topic. This work has been supported, in part, by
NSF, NASA and the DOE at the University of Chicago,
by the DOE and by NASA through grant NAGW 2381 at Fermilab
and by the Alexander-von-Humboldt foundation.
\bigskip

{\it Note added:} During preparation of this manuscript we
became aware of a paper by P.~Sommers~\cite{Sommers} which essentially
deals with the same topic. The conclusions reached concerning the
shock acceleration scenario are quite similar. In difference,
however, the aim of our paper was a more detailed discussion
of all possible explanations related to relativistic particles and
especially to note the emergence of the TD scenario as an interesting
option.

\newpage

\begin{table}[h]
\renewcommand{\arraystretch}{1.5}
\begin{tabular}{|l|c|c|c|}\hline
  \multicolumn{4}{|c|}{protons and secondary $\gamma$'s and $\nu$'s}\\ \hline
   & \multicolumn{2}{c|}{maximal energy} & additional \\
  Sources & shock acc. & drift acc. & problems \\ \hline
  AGNs & $10^{20}\eV\left({RB\over10^3\kpc\muG}\right)$ &
  $10^{21}\eV\left({RB\over10^3\kpc\muG}\right)$ & distance+direction \\ \hline
  pulsars & $2\times10^{17}\eV\left({RB\over0.002\pc\G}\right)$ &
  $2\times10^{18}\eV\left({RB\over0.002\pc\G}\right)$ & direction \\ \hline
  galactic wind & \multicolumn{2}{c|}{$3\times10^{17}\eV
  \left({u\over600\km\sec^{-1}}\right)^2
  \left({T_A\over1.5\times10^{10}{\rm y}}\right)\left({B\over0.1\muG}\right)$}
  & \\ \hline\hline
  \multicolumn{4}{|c|}{galactic heavy nuclei}\\ \hline
   & \multicolumn{2}{c|}{maximal energy} & additional \\
  Sources & shock acc. & drift acc. & problems \\ \hline
  pulsars & $2\times10^{17}\eV\,Z\left({RB\over0.002\pc\G}\right)$ &
  $2\times10^{18}\eV\,Z\left({RB\over0.002\pc\G}\right)$ & direction \\
  \hline
  galactic wind & \multicolumn{2}{c|}{$3\times10^{17}\eV\,Z
  \left({u\over600\km\sec^{-1}}\right)^2
  \left({T_A\over1.5\times10^{10}{\rm y}}\right)\left({B\over0.1\muG}\right)$}
  & \\ \hline\hline
  \multicolumn{4}{|c|}{mainly $\gamma$'s, some protons and $\nu$'s}\\ \hline
   & \multicolumn{2}{c|}{maximal energy} & problem\\ \hline
  TD's & \multicolumn{2}{c|}{$\ga10^{24}\eV$} & exotic\\ \hline
\end{tabular}
\caption[1]{Maximal energies and problems with the options discussed in
this paper.}
\end{table}

\newpage
\centerline{\bf Figure Captions}

\bigskip

\noindent{\bf Figure 1A and 1B:} Distance of source versus source energy
for protons of arrival energies of $E_0=1.7\times10^{20}\eV$ and
$3.2\times10^{20}\eV$, respectively. Plotted from bottom to top are the
average distances and the maximal distances at the $1\sigma$, $2\sigma$
and $3\sigma$ level, respectively.

\bigskip

\noindent{\bf Figure 2:} Arrival directions of the highest energy
events seen by Fly's Eye and the Yakutsk experiment in galactic
coordinates~\cite{Covault}. This plot is centered
around the galactic anticenter with the middle horizontal line being
the projection of the galactic plane.
Also shown are nearby galaxy clusters (big circles), AGN's (small
circles) and galactic supernova remnants (light circles). As discussed
in section 3.3 the galactic magnetic field is supposed to have a
coherent component near the galactic plane which in the outer region
is polarized in the direction of decreasing galactic longitude. Therefore,
the apparent arrival direction of cosmic rays coming from one
of the objects shown should be shifted by $\sim1.5^o\,Z(3\times10^{20}\eV/E)$
to lower latitude.

\end{document}